%% file: main.tex
\documentclass[sigconf]{acmart}
\renewcommand\footnotetextcopyrightpermission[1]{} % removes footnote with conference information in first column
\usepackage{booktabs} % For formal tables

\usepackage{balance}
\begin{document}
\title{Personalizing Similar Product Recommendations in Fashion E-commerce}
% \titlenote{Produces the permission block, and
% \subtitle{Extended Abstract}
% \subtitlenote{The full version of the author's guide is available as
%   \texttt{acmart.pdf} document}

\author{Pankaj Agarwal}
\affiliation{%
  \institution{Myntra Designs, India}
}
\email{pankaj.agarwal@myntra.com}

\author{Sreekanth Vempati}
\affiliation{%
  \institution{Myntra Designs, India}
}
\email{sreekanth.vempati@myntra.com}

\author{Sumit Borar}
\affiliation{%
  \institution{Myntra Designs, India}
}
\email{sumit.borar@myntra.com}

% \renewcommand{\shortauthors}{B. Trovato et al.}

\begin{abstract}
In fashion e-commerce platforms, product discovery is one of the key components of a good user experience. There are numerous ways using which people find the products they desire. Similar product recommendations is one of the popular modes using which users find products that resonate with their intent. Generally these recommendations are not personalized to a specific user. Traditionally, collaborative filtering based approaches have been popular in the literature for recommending non-personalized products given a query product. Also, there has been focus on personalizing the product listing for a given user. In this paper, we marry these approaches so that users will be recommended with personalized similar products.
Our experimental results on a large fashion e-commerce platform (Myntra) show that we can improve the key metrics by applying personalization on similar product recommendations. 
\end{abstract}

\maketitle

\input{personalized_similar}

\bibliographystyle{ACM-Reference-Format}
\bibliography{main}

\end{document}

%% file: personalized_similar.tex
\section{Introduction}

% Why are we solving this problem? 
In e-commerce, the number of products in the shelf space are practically infinite. Thus, the users have to navigate through a plethora of options in any category before making a purchase and they often get disinterested in the process soon. This problem is more prominent in fashion compared to other e-commerce domains like that of movies, books, electronics, etc. In other categories, the users generally have a crisp understanding of what they want to buy. In fashion, the users mostly don't know what they want. Even if they know, it is hard for them to explain it to the search engine which understands a product with limited taxonomy of attributes. Limited real estate in mobile screens aggravates the problem further. The faster we can assist a user in finding the right product, higher are the chances of user conversion. Hence, personalization becomes an important lever to cater to diverse users' need, allowing for better product discovery and customer experience. In this paper, we propose an approach to personalize similar products that are being shown to the user for a given product. We perform our experiments on data collected from Myntra, which is a large e-commerce platform in India and show how our approach performs better compared to non-personalized recommendations. 

\begin{figure}[t]
    \includegraphics[width=60mm]{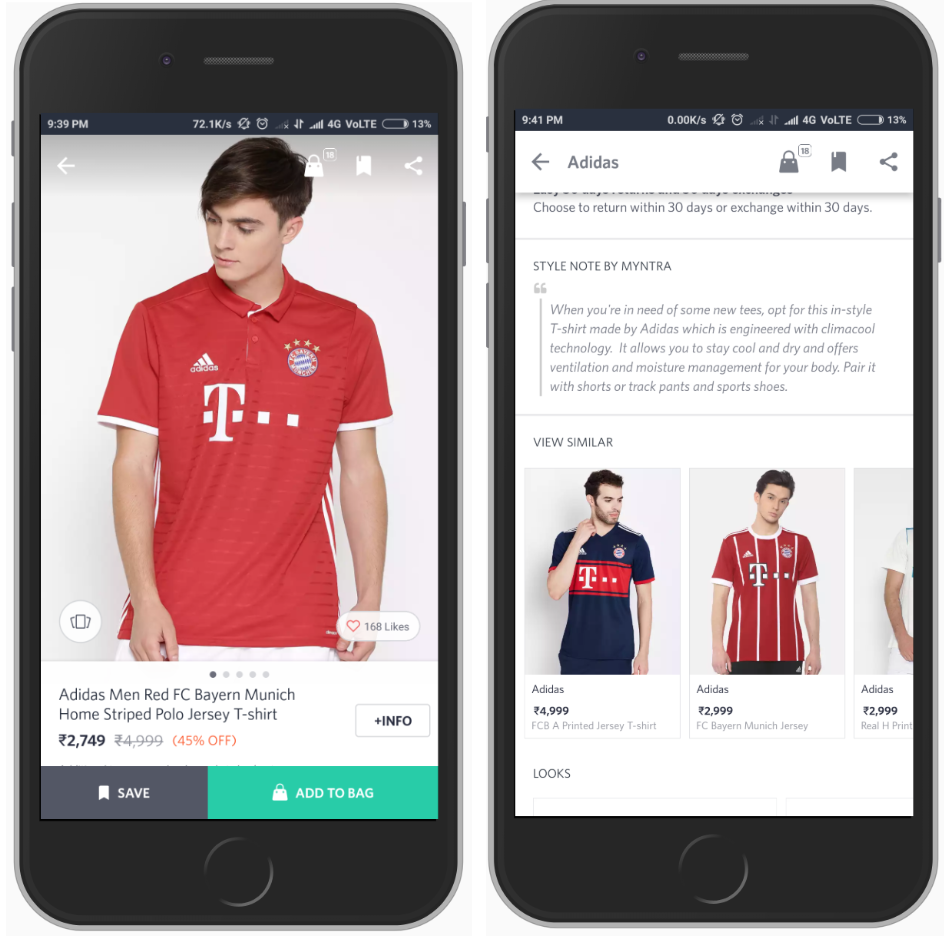}
\caption{Example screenshots of Myntra mobile app showing how similar products are typically displayed to the user on our platform.}
    \label{fig:similar}
    \vspace{-2mm}
\end{figure}

 Similar Products is a great way to recommend certain products to users based on current context \cite{amazon}. These recommendations are highly useful for a user if he/she has liked a certain kind of product and may buy if presented with few more slightly varied products. Figure \ref{fig:similar} shows how the similar products are displayed to the user for a given query product on our platform.
 
Typically, similar product recommendations are solved through either content based or collaborative filtering based approaches. Content based filtering approaches recommend products by using the attributes of the products. Collaborative filtering approaches use historical user product interactions. In fashion, products can be represented with their product attributes like colour, pattern, fabric, sleeve type, collar type etc. These attributes are seldom fixed and usually change with new trends. Another challenge is to tag attributes for all the products manually. Further even after tagging, users' taste is often complex and are hard to explain in terms of the limited set of attributes. Thus collaborative filtering based approaches are preferred over content based ones. 

Typically, algorithms for similar products recommend a non-personalized set of products to all the users i.e. the result set is completely agnostic of the user \cite{amber_paper}\cite{amazon}. Though the results are derived considering the browsing behavior of all the users, the recommended results tend to favour the choices of majority of the population while ignoring an individual's subtle preferences. Figure \ref{fig:personalization_example} depicts the general recommendations against a query product which is an orange color solid shirt dress. Non - personalized recommendations are shown in first row. However, the recommendations can be re-ranked if we have certain information about the user. Below are the two possible examples:

\begin{figure*}
    \includegraphics[width=160mm]{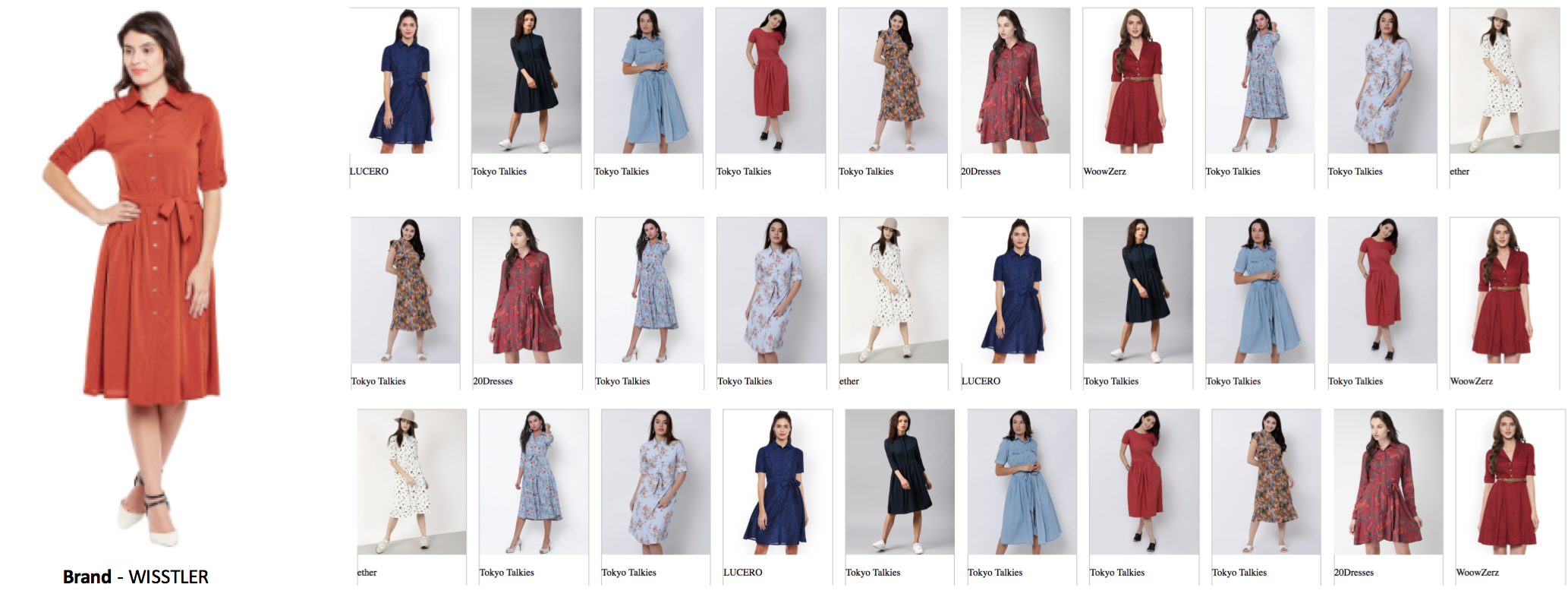}
    \caption{Through the above image, we illustrate how user level personalization can improve similar product recommendations. On the left hand side, we have a query product. On the right hand side, the first row shows the non-personalized similar product recommendations. The second row shows how ideal ranking will look like if the user generally likes floral dresses. And the third row shows the ranking in case of a user who has affinity towards lighter colours }
    \label{fig:personalization_example}
    \vspace{-2mm}
\end{figure*}

\begin{itemize}
    \item Lets say we know that the user has strong affinity towards floral pattern compared to solid. Then, if we can recommend floral styled shirt dresses to the user on top, it should result in better recommendations.  This is depicted in 2nd row of Figure \ref{fig:personalization_example}. 
    \item If we know that the user has strong affinity towards light colours then it makes sense to recommend the white dress as a top result. This is depicted in 3rd row of Figure \ref{fig:personalization_example}. 
\end{itemize}

In literature, we find solutions for product recommendations with query being either a product or a user\cite{bpr}\cite{amazon}\cite{recommendations_book}.In this paper, we propose an approach to solve the problem of personalizing similar products with query as both user and a product.
	 
We had to overcome few challenges to incorporate user's taste into the system. Firstly, our platform i.e Myntra has about 6 million products available at any given time \& the largest category which is T-shirts for Men has about 50k items. Further the data suffers from huge long tail because of which the interaction signals are sparse. For instance, a typical user-item matrix on our platform would have a sparsity of <0.1\%. On our platform, 20\% of products lead to more than 80\% of  revenue on a daily basis. Secondly, it is very hard to find out a user's affinity towards all the possible attributes in a particular category. It would not be succinct if we try and represent user's taste with few commonly known attributes \cite{community}.

Our approach combines the solutions of finding similar product recommendations and user level personalization. We use matrix factorization based approaches for this purpose and thereby overcome the challenges mentioned above.

% The MF algorithm that we used in the paper takes care of the sparsity problem as well as representing a user's taste in 'K' latent dimensions.  

In the following sections of the work, we describe approaches to solve the problem and discuss few experiments that show  how personalizing the similar product recommendations improves key metrics.

\section{Related Work}
Our work is related to two areas: recommendation systems and personalization systems. There has been significant work done already on recommendations systems in various domains \cite{recommendations_book} like ecommerce \cite{amazon}, news \cite{google_news} and  music \cite{music_spotify}.

Collaborative filtering based systems have been very popular for recommendations \cite{amazon} \cite{cf_implicit} \cite{bpr} \cite{matrix_factorization}. In \cite{google_news}, a large scale collaborative filtering based system is proposed for personalizing news to a given user. In \cite{music_spotify}, a deep content based music recommendation system is proposed to tackle the lack of user interactions data. 

Further improvements to the recommendation algorithms were also presented in the \cite{visual_bpr}, \cite{dvbpr}, \cite{xdeepfm}, \cite{wide_deep}. Our work is focused on using these approaches for personalizing the similar product recommendations.

Context driven recommendations systems have shown to improve the existing performance of recommendations in \cite{context_based}. In \cite{context_aware_fm}, authors propose a way of incorporating context into the recommendation systems specifically on how, when and why a rating was done by a user. There are also a set of works which solve the cold-start problem, for example in \cite{visual_bpr} visual features are used.

In \cite{youtube}, a deep learning based video recommendation system is proposed which marries personalization with recommendation and is one of the closest work we have followed in terms of the objective. In \cite{personalization_fashion}, it is shown the personalization is one of the prominent factors effecting key metrics in online shopping.

Note that while our earlier work \cite{amber_paper} focuses on non-personalized similar product recommendations, this paper's primary focus is on personalized similar product recommendations.

\begin{figure*}
    \centering
    \includegraphics[width=180mm,scale=0.5]{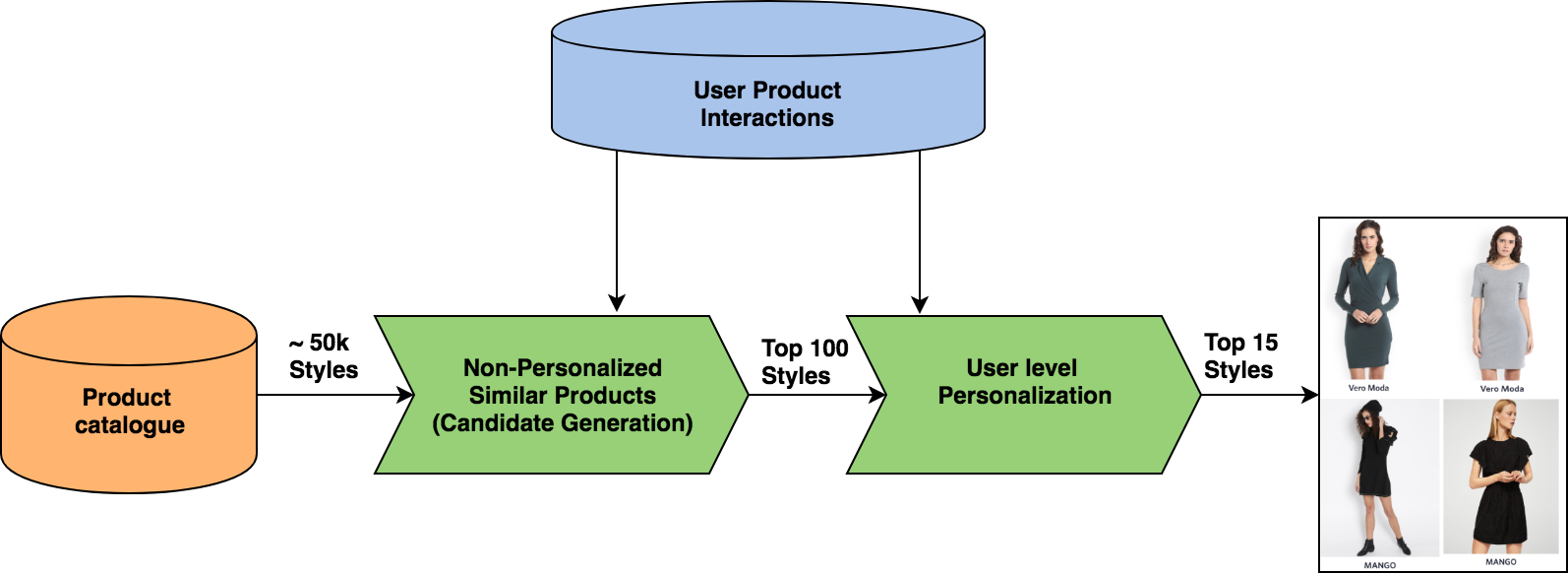}
    \caption{Overview of the approach. We first generate a candidate set using non-personalized similar product generation algorithm and then refine the results using user preferences.}
    \label{fig:approach}
\end{figure*}

\section{Methodology}\label{methodology}
Given a user $u$ and current product $P_i$ which is being viewed by the user, our objective is to come up with an affinity score denoted by $f(P_j|P_i, u)$ for each product $P_j$ in the catalogue. This score will be used to rank and display the products to the user. We compute this score as a linear combination of the following:

\begin{itemize}
\item $h(P_j,P_i)$ representing the similarity of product $P_j$ with the given product $P_i$.
\item $g(P_j,u)$ representing the similarity of product with the user's taste.
\end{itemize}

Figure \ref{fig:approach} summarizes our approach. We first explain the input data that is being collected and then we describe how the above two scores are calculated separately and then combined to generate final ranking.

\textbf{Interactions data:} In the absence of explicit ratings for products by users on our platform, we depend on implicit signals from the users. These implicit signals include product list\_views (number of times a product is seen by the user in the product search listing), clicks (number of times a user has viewed the product details page), add to carts \& orders. We assign a rating for a product by the user as a weighted sum of these signals. We consider these weights also as hyperparameters for our approach. Using this we form a user interactions matrix, where each row corresponds to a user and each column corresponds to a product. We use this data as an input for our approach.

For new products and new users, we use the content based methods for recommending similar products and don't personalize for them. 

\subsection{Non-personalized similar products \textit(Candidate Generation)}
For generating non-personalized similar recommendations, we use collaborative filtering based approach. In our earlier work \cite{amber_paper}, we have already shown how collaborative filtering approaches perform better compared to content based approaches.
%\footnote{published in a top conference, not revealed to preserve anonymity},
Further, we have experimented with item-item collaborative filtering approach and matrix factorization based approaches \cite{bpr}\cite{cf_implicit}\cite{matrix_factorization}. From our experiments, we found that item-item collaborative filtering approach performs better on our dataset. 

In item-item collaborative filtering approach, we use the vectors corresponding to each product from the user-item interactions matrix. The similarity between given two products is then calculated as the cosine similarity between the vectors. From the sorted set, we choose the top results which acts as our candidate set for the next step. Choosing the top results helps in faster response times for production systems. 
% If $x$ and $y$ represent two product vectors, then cosine similarity between those two products is calculated as follows:
% \begin{equation}
%     cos(\pmb x, \pmb y) = \frac {\pmb x \cdot \pmb y}{||\pmb x|| \cdot ||\pmb y||}
% \end{equation}

\subsection{User level personalization}
For personalization, we have considered two popular matrix factorization based approaches specifically Implicit Alternating Least Squares (ALS) \cite{cf_implicit} and Bayesian Personalized Ranking(BPR) approaches \cite{bpr}. Note that these results can be further improved by using other sophisticated approaches.

These algorithms work by transforming the sparse user interaction matrix into low dimensional latent space vectors for both the users and products. The transformed vectors represent each user and product with their low dimensional dense vectors. The user vector captures the user's fashion taste in latent space and product vectors captures the hidden attributes in the same space. We briefly explain both of these approaches below:

\textbf{Implicit Alternating Least Squares (ALS-MF):}
This algorithm \cite{cf_implicit} is designed to work on implicit ratings and optimizes the modified cost function compared to traditional MF approaches\cite{matrix_factorization}. Cost function for this method is written as:

\begin{equation}
    min \sum_{u,i}{}{c_{ui}(p_{ui} - x^T_u y_i )^2 + \lambda(\sum_{u}||x_u||^2 + \sum_{i}||y_i||^2) }
\end{equation}

In the above equation, $x_u$ represents the latent user vector and $y_i$ represents the latent product vector in $k$ dimensions. The preference of the user $u$ for the product $i$ is given by $x_u^T y_i$. 
$p_{ui}$ represents observed preference score obtained from the implicit signals for the user $u$ and product $i$. And, $c_{ui}$ represents the confidence on the implicit signals and $\lambda$ is the regularization parameter. $c_{ui}$ and $\lambda$ are hyper-parameters and their exact values are determined by cross-validation.

Cost function tries to minimize the difference between the estimated score and observed score across all the user and product combinations. 

\textbf{Bayesian Personalized Ranking (BPR-MF):}
In the ALS approach the focus is on estimating the point wise score correctly, whereas BPR \cite{bpr} works on optimizing the pairwise ranking of the products for a user correctly. For this purpose, the model optimizes the loss function which considers pair of products for each user. The loss function for BPR in general is written as:

\begin{equation}
   -\sum_{(u,i,j)}{}{\ln \sigma(x_{uij}) + \lambda_\Theta || \Theta ||^2}
\end{equation}

where $u,i,j$ are the triplets of user $u$ and product pairs $(i,j)$ available in the interactions dataset such that user likes product $i$ over product $j$. And, $x_{uij}$ denotes the difference of estimated preference scores for the user $u$ to the product $i$ and product $j$. $\Theta$ is the model parameter vector and $\lambda_\Theta$ are model specific regularization parameters. In the case of matrix factorization, the model parameters $\Theta$ are user and item vectors. 

We feed the user interactions data to the above explained BPR-MF and ALS-MF approaches to generate the user and product latent vectors. Using these vectors we compute a score for each user and product. The score represents the affinity for the user towards the product. We refer this as user-product similarity score. The sorted list based on these scores gives us a personalized product listing for a given set of products.

% We used variants of the MF approach to find dense representations of users as well as products in the same space. We can thus compute cosine distances between User-User, User-Product \& Product-Product entities. 
% The User-Product cosine distance represents the affinity of a user towards a particular product.
% <Write a bit about the ALS  \& BPR approaches>

\subsection{Personalized similar products (Final Ranking)}\label{method}
Once we obtain the non-personalized similar products set, we need to incorporate personalization to re-rank these products. One of the ways to get a personalized product listing is to directly use the user-product similarity scores assuming that all the products are equally similar to the query product. But this is seldom the case \& thus we used a combination of product-product similarity scores and user-product similarity scores so that we can ensure that the resulting list preserves the context which is the query product as well as the effect of personalization by preserving the user's taste. We combine the scores as follows:
\begin{equation}
f(P_j|u, P_i) = \alpha h(P_j, u) + (1-\alpha) g(P_j,P_i)
\end{equation}

where $\alpha$ is a hyper-parameter which is determined using cross-validation.
	
\section{Experiments \& Results}
\subsection{Dataset}
For all our experiments, we use the clickstream and purchase data of users. We split our user interactions data into two non overlapping sets, a training set, which is used to generate the user and product vectors in latent space, and a test set, which is used to evaluate the approaches. Training set consists of data for $12$ months and test set consists of data from next $1$ month. The training set consists of \textasciitilde $50k$ products with \textasciitilde $1.2M$ unique users. Sparsity of the interactions matrix is $0.1\%$. Note that the users and products will be common in both the training and test set. We present our experimental results for the category ``Men-Tshirts". The results were seen to be consistent across all the other categories.

\subsection{Evaluation Metric}

As this is a ranking problem, we have used standard Information Retrieval metrics namely  Precision, Recall and Mean Average Precision$@$K. The value of "K" is chosen to be 15 in our case since the number of recommendations we would show to the end user is the same. Precision and recall are single value metrics and are used to gauge the performance of all the results together irrespective of the order. However, order of results matter in information retrieval \& mean average precision (MAP) takes that into account.
For computing MAP, the predicted result set is the top 15 products recommended by our algorithm for a given query product.
Ground truth set is obtained from the test set. Query product is chosen as the first product a user interacted with \& remaining products are assigned to the ground truth set. 

% Average Precision is defined as
% 	\begin{equation}
% 	    AveP = \sum\limits_{i=1}^K P(i)*\Delta R(i) 
% 	\end{equation}
% where $P$ is the precision and $R$ is the recall.

\subsection{Computing the rating function}

As explained earlier, we don't have explicit ratings on our platform. So, we have used clickstream data to come up with a rating for a product and a user. The rating is computed as a weighted linear combination of the following quantities : list\_views, clicks, add\_to\_carts and purchases. All the events can provide different weights to the rating computation. The frequency of the event or number of times a user performed an event (say clicked on product display page ) can also be factored into the rating function. To understand the change in evaluation metric based on varying weights we performed a grid search. The Table 1 reflects few examples from the results. We conclude that weightage set of (0.25,1,1,1) provides us with the best MAP. The frequency of event was found to be not useful for the model \& hence we ignore the frequency from our computation for the rest of the paper.

\subsection{Non Personalized Similar Products (Candidate Generation)}
To generate similar products (Candidate set), we have done experiments using Item-Item collaborative Filtering, ALS-MF \& BPR-MF. From the resultant product vectors, we computed cosine distance between all the products. For each product, we sort based on these scores to get the similar products ranking. We have found that Item-Item collaborative filtering performs best for this purpose. We choose top 100 products against a query product as our candidate set.

\subsection{Algorithm and Confidence Optimization}
\begin{figure}[]
    \centering
    \includegraphics[width=80mm]{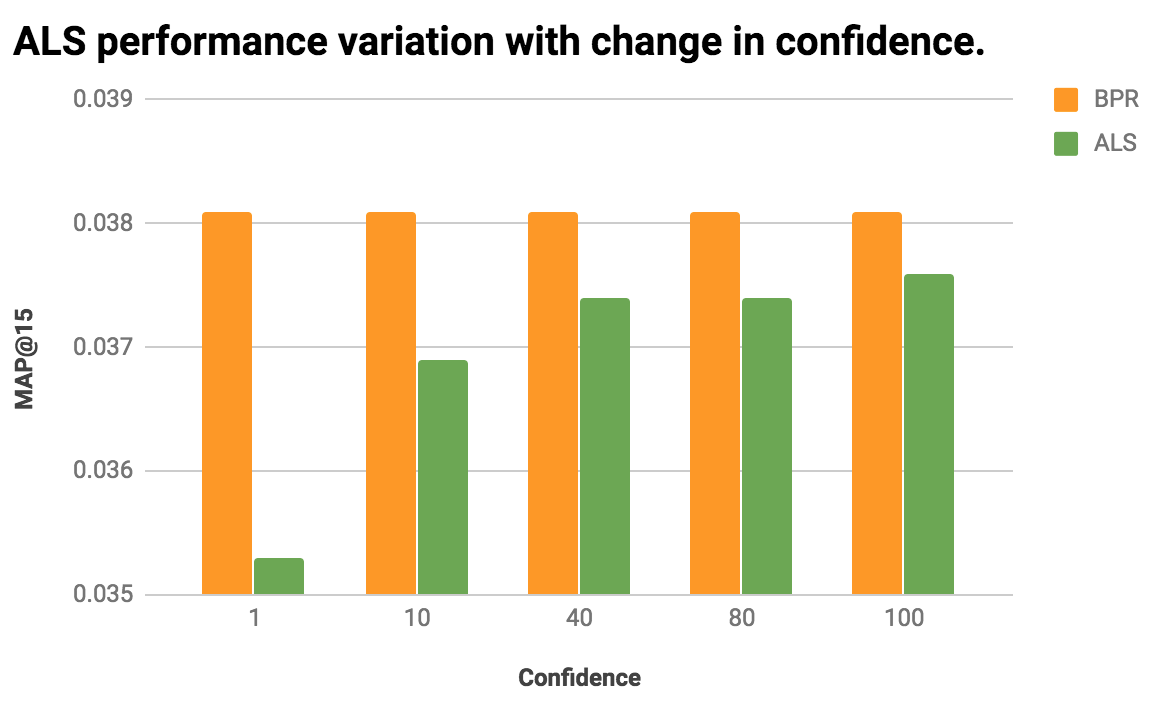}
    \caption{Performance variation with change in confidence parameter for both ALS-MF. We also plot the MAP@K using BPR-MF.}
    \label{fig:algo_comparison}
    %\vspace{-4mm}
\end{figure}

For ALS-MF, confidence parameter determines the weightage of the implicit rating. In the table \ref{table:rating1}, we show how the weightage of different implicit signals impacts MAP value.  Figure \ref{fig:algo_comparison} shows the variation in MAP with varying confidence and choice of algorithm. Based on the results we decided to stick with BPR-MF algorithm. In our implementations, we have used map-reduce framework for the large scale data processing and \cite{implicit_code} for implementing the ALS-MF and BPR-MF algorithms.

\subsection{Finding personalized similar products (Final Ranking) }
\begin{figure}[]
    \centering
    \includegraphics[width=80mm]{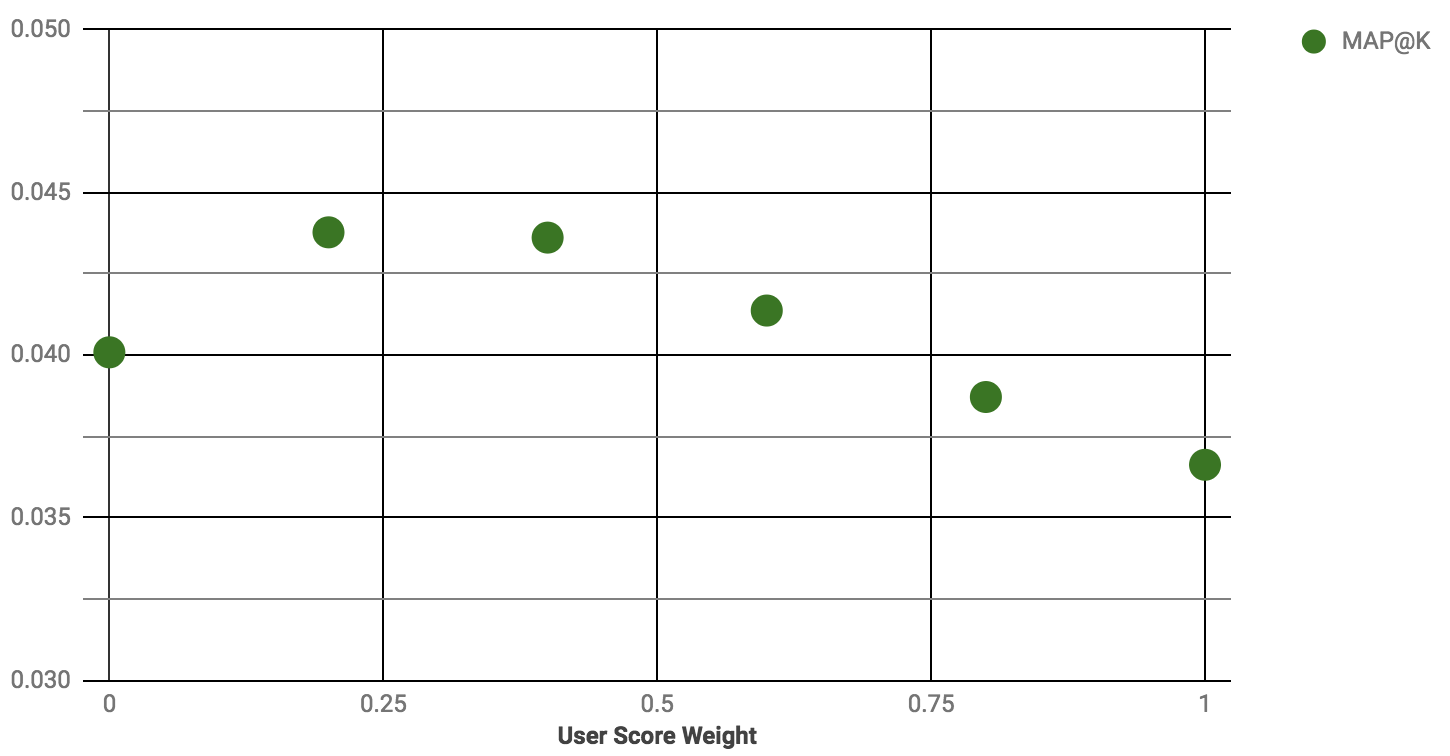}
    \caption{Graph shows how the performance varies with change in weight of user level score.}
    \label{fig:user_weight}
    \vspace{-4mm}
\end{figure}

The objective was to ensure that the results obtained are relevant for the current context as well as well suited to the taste of the user. As explained in the section\ref{method}, we use a linear combination of non-personalized similarity scores and user level scores. We have done experiments with $\alpha$ as hyper-parameter. In the Figure \ref{fig:user_weight}, we show the effect of change in $\alpha$. We can see that providing $80 \%$ weightage to the similar products and $20 \%$ weight to user preferences gives with the optimal MAP$@$K.

\begin{table}[]
\centering
\begin{tabular}{|l|l|l|l|l|l|}
\hline
\textbf{ListViews} & \textbf{Clicks} & \textbf{Carts} & \textbf{Orders} & \textbf{Freq} & \textbf{MAP@15} \\ \hline
0             & 1            & 1             & 1              & 0                  & 0.0432              \\ \hline
0             & 1            & 1             & 1              & 1                  & 0.0378              \\ \hline
\textbf{0.25}          & \textbf{1}            & \textbf{1}             & \textbf{1}              & \textbf{0}                  & \textbf{0.0437}              \\ \hline
0.25          & 1            & 1             & 1              & 1                  & 0.0393              \\ \hline
0.25          & 10           & 4             & 1              & 0                  & 0.0296              \\ \hline
0.25          & 10           & 4             & 1              & 1                  & 0.0295              \\ \hline
\end{tabular}
\caption{Performance variation by changing weightages of different implicit signals using BPR-MF. We show few data points in this table. Highlighted row works the best.}
\label{table:rating1}
\end{table}

\vspace{4mm}

\section{Conclusion}
We have presented a method to personalize the similar product recommendations. We have shown how our approach improves the mean average precision metric on a large dataset collected from our e-commerce platform Myntra. Further, we will be deploying this solution in production and validate this by performing A/B test. As future work, we plan to combine this with approaches which use visual features. 

\section{ACKNOWLEDGMENTS}
The authors would like to thank Sabbarish R, Ankul Batra, Sagar Arora and Ghani Mohammed Abdulla for their contributions in reviewing this work and for providing valuable inputs to algorithm design, implementation and evaluation. 

% The authors would like to thank Jim McFadden and Pranav Khaitan for valuable guidance and support. Sujeet Bansal,Shripad Thite and Radek Vingralek implemented key components of the training and serving infrastructure. Chris Berg and Trevor Walker contributed thoughtful discussion and detailed feedback.

\balance